# Fragility and molar volumes of non-stoichiometric chalcogenides – the crucial role of melt/glass homogenization

**R. Bhageria[1], K. Gunasekera[1], P. Boolchand[*,1] and M. Micoulaut[2]**

[1] Department of Electrical and Computing Systems, College of Engineering and Applied Science, University of Cincinnati, Cincinnati OH 45221-0030. USA.

[2] Laboratoire de Physique Théorique de la Matière Condensée, Université Pierre et Marie Curie, Boite 121, 4, Place Jussieu, 75252 Paris Cedex 05, France.

Melt-fragility index (m) and glass molar volumes ($V_m$) of binary Ge-Se melts/glasses are found to change reproducibly as they are homogenized. Variance of $V_m$ decreases as glasses homogenize, and the mean value of $V_m$ increases to saturate at values characteristic of homogeneous glasses. Variance in fragility index of melts also decreases as they are homogenized, and the mean value of m decreases to acquire values characteristic of homogeneous melts. Broad consequences of these observations on physical behavior of chalcogenides melts/glasses are commented upon. The intrinsically slow kinetics of melt homogenization derives from high viscosity of select super-strong melt compositions in the Intermediate Phase that serve to bottleneck atomic diffusion at high temperatures.

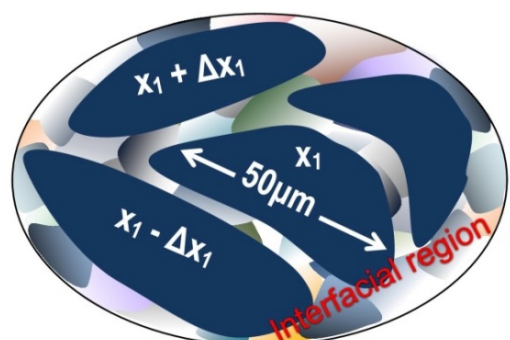

Schematic of a near homogeneous $Ge_xSe_{100-x}$ melt composed of homogeneous (dark) regions separated by heterogeneous (light) interfacial ones.

## 1 Introduction

Mass density of solids contains direct information on their atomic packing. Density can be measured rather accurately using the age old Archimedes' principle that was enunciated in about 250 BC Syracuse, Italy. In the case of the network glasses, $V_m$ acquires fundamental importance largely because glassy solids like proteins form space filling networks[1]. Glasses possess densities that are typically 90% of their crystalline counterparts. For example, vitreous silica (density 2.20 gms/cm$^3$) has a somewhat smaller density than its high T crystalline form cristobalite (2.33gms/cm$^3$) and tridymite (2.28 gms/cm$^3$)[2]. Variations of $V_m(x)$ in network glasses often display global minima in select compositional windows that are characteristic of Intermediate phases[3-5], a feature of compacted glasses[6] For example, in the $Ge_xSe_{100-x}$ binary glasses of proven homogeneity[7, 8] (Fig.1), one finds a broad global minimum of $V_m(x)$ in the Ge concentration range, 19.5% < x < 26%. On either side of this minimum, $V_m(x)$ increases rather steeply in homogeneous glasses but less so in heterogeneous ones. For networks that are isostatically rigid, i.e., have the optimal[9] counts of bond-bending and bond-stretching forces per atom of 3, long range Coulombic and Van der Waals forces assist in compacting networks. The broad minimum of $V_m$ results generally due to such stress-free or optimally coordinated networks adapting to expel the stress creating redundant bonds. Window glass is an example of a compacted glass[10]. Recently, it was shown that room temperature relaxation of Gorilla glass[11] is strictly volumetric in nature, i.e., determined by long range forces. Compacted networks have been observed in chalcogenides[3, 17, 18], modified oxides[5, 19], heavy metal oxides[20]. Remarkably, proteins also form compacted networks, and display many features of self-organization as observed in glassy networks[21].

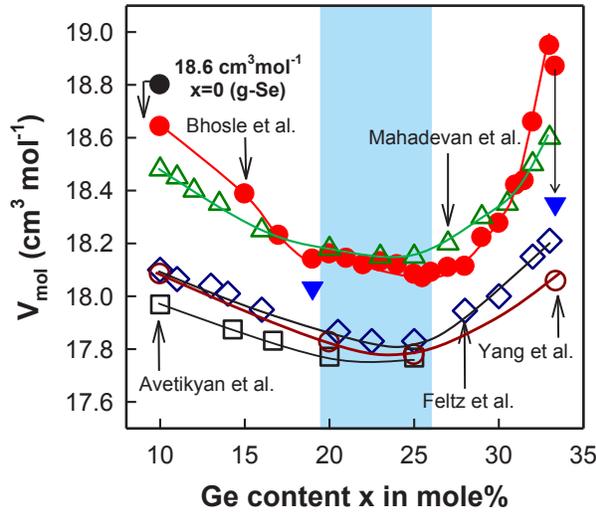

**Figure 1** Molar volumes $V_m(x)$ of dry and homogeneous $Ge_x$-$Se_{100-x}$ glasses (●) from Bhosle et al[12]. compared to other reports in the literature. Mahadevan et al. ref [13] (Δ), Feltz et al. ref [14] (□), Avetikyan et al ref [15] (□) and Yang et al. ref [16] (○). The two (▼) data points are of wet samples. In the shaded panel, $V_m$ shows a global minimum, a compositional window which corresponds to the Intermediate phase of the present binary. The $V_m$ at $x = 0$ for pure g-Se is also included from our work.

In the $Ge_xSe_{100-x}$ binary there have been several earlier reports of molar volumes, $V_m(x)$, across a wide range of Se-rich ($x < 33.3\%$) compositions. Some of these reports (Fig.1) reveal a measurably lower value of $V_m$ (Fig.1) than the ones reported by Bhosle et al.[12] in glasses of proven homogeneity. These differences in $V_m$ are much too large to be due to uncertainty of density measurements or due to cooling rate effects, which affect volume by a fraction of a percent[22, 23]. Here we visit the issue and show that the consistently low values of $V_m$ across the wide range of compositions reported by several groups, most likely, stem from heterogeneity of melts/glasses by virtue of synthesis. Discrepancies in other reported physical properties, such as fragility for example, also stem from lack of melt homogeneity.

Viscosity measurements as a function of temperature on stoichiometric chalcogenides, oxides, sugars and alcohols have been widely used to understanding dynamics of supercooled melts. Since viscosity η is proportional to τ, these viscosity measurements essentially probe shear relaxation time τ with temperature. If one plots $log(\tau)$ against $T_g/T$, the dimensionless slope, m, near $T = T_g$ defines the fragility index-m.

$$m = \lim_{T \to T_g} \left| \frac{dlog\tau}{d(T_g/T)} \right| \quad (1)$$

Melts possessing a high (low) value of m are defined to be strong (fragile), and are found to display a strongly non-exponential (Arrhenian) variation of the relaxation time τ(T). In non-stoichiometric chalcogenides glasses experiments reveal that fragility index can vary non-monotonically with composition displaying a fragile to strong variation. For example, in the $(Ge_{1/2}As_{1/2})_{100-y}Se_y$ ternary, Tatsumisago et al.[24] found a broad minimum of the fragility index m near r = 2.40 the rigidity percolation threshold. Here r represents the mean coordination number.

Fragility index (m) of specially homogenized $Ge_x$-$Se_{100-x}$ melts were reported recently from complex $C_p$ measurements[4], and one found (Fig.2) that in the composition range $20\% < x < 26\%$, or mean coordination number range $2.40 < r < 2.62$, m became quite low, i.e., m < 20. Furthermore, in the narrow composition range, $21.5\% < x < 23.0\%$, m acquired a specially low value of ~15.0, i.e., melts became super-strong. By directly mapping melt stoichiometry during melt-reaction/equilibration at high temperatures (950°C), one also showed[4] that the super-strong behavior of melts in that narrow composition range serves as a bottleneck to batch homogenization. These fragility data underscore a close connection between fragility and network topology as also demonstrated theoretically from a harmonic oscillator model reproducing the radial and angular forces constraining a network at a molecular level25. As networks self-organize and adapt under increasing stress/Ge composition, melts become strong, and near the center of the IP, melts actually become super-strong. Clearly, the low fragility value of such compacted networks is apparently connected to the existence of extended range structural correlations in such melt compositions corresponding to the IP.

**2 Raman profiling as a method to synthesize melts of controlled heterogeneity** Melt-quenching as a method to synthesizing glasses consists of homogeneously alloying element A with B in a suitable ambient, and then supercooling the melt once it is homogeneous, to bypass crystallization and realize a bulk glass. It is important to emphasize that unless the melt is homogeneous, one cannot expect the glass to be so. Recently, we introduced a novel method to establish heterogeneity of quenched melts by recording Raman scattering along the length of a melt column non-invasively[12], and found the method works remarkably well for chalcogenides. In these Raman profiling experiments one observes spectacularly different Raman lineshapes (Fig.3) along the length of a melt column in the early stages (less than 2 days) of reaction when melts are heterogeneous, but these differences steadily disappear as all spectra coalesce into one unique spectrum upon prolonged reaction (> 7 days) as melts homogenize. We illustrate the observation for a melt at a composition x = 10% in the present binary in Fig.3 that we acquired in the present work. Similar Raman data were acquired at x = 15%, a composition for which reaction of the elements took much longer (17days) for the batch to homogenize.

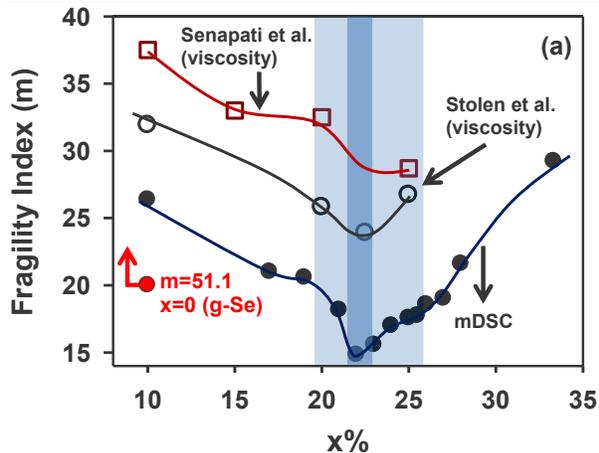

**Figure 2** Fragility index of binary $Ge_xSe_{100-x}$ melts reported in viscosity and complex Cp measurements. Figure taken from ref.[4] The light grey region represents the Intermediate Phase. The dark grey region the fragility window. The viscosity derived fragility from the work of Senapati et al.[26] and Stolen et al.[27] are included. Fragility of pure Se glass from our mDSC measurements is indicated.

The Raman profiling method also makes accessible melt-quenched glasses of *varying heterogeneity* by merely tuning the duration of the alloying process in days(d) in the $1d < t_R < 16d$ range. In this work we have exploited that capability, and have examined the variation of fragility index (m), and molar volume ($V_m$) of $Ge_xSe_{100-x}$ melts/glasses at specific compositions x as a function of their *heterogeneity*.

**3 Experimental** Bulk $Ge_xSe_{100-x}$ glasses were synthesized by sealing 2 gram sized batches of 99.999% Ge and Se in evacuated ($2 \times 10^{-7}$ Torr) quartz ampoules and reacted at 950°C for periods up to 17 days keeping ampoules vertical in a T-regulated box furnace. Details appear in ref[7,8]. Melts/glasses were synthesized at x = 10% and at x = 15%. Four identical ampoules were sealed and reacted respectively for $t_R$ = 1d, 2d, 4d, 6d. The x = 15% sample needed to be reacted up to 17d to homogenize. After such periods melts were quenched the usual way, and FT-Raman profiles acquired using a Thermo-Nicolet model Nexus 870 system. Quenched melts were aged at room temperature for several days under the same conditions, prior to undertaking molar volume measurements. Next quartz tubes were opened and mass density of the glasses measured using a digital microbalance model B154 from Mettler Toledo. In a typical measurement 150 to 200 mg pieces were placed on a hooked quartz fiber suspended from the balance pan and their weight measured in air and then in 200 Proof Ethyl alcohol. We made efforts to measure at least 5 samples from a given batch composition to sample nearly 50% of the batch composition, and establish the *variance in density* across the batch. The alcohol density was calibrated using a Si single crystal. (ρ = 2.329 gm/cc). And the accuracy of the density measurements independently checked by measuring the density of a Ge single crystal (ρ=5.323 gm/cc). Melt fragility index were established by examining the complex Specific heat as a function of modulation frequency using a Q2000 Modulated DSC from TA Instruments. Details of the measurements are discussed elsewhere[28]. The in-phase and out-of-phase components of the complex specific heat, $C_p$, were measured as a function of modulation frequency. The in-phase $C_p$ shows a rounded step, while the out-of phase $C_p$ a Gaussian-like peak. At the peak, the condition $\omega\tau = 1$ is fulfilled, i.e., the melt completely relaxes to follow the programmed modulated heat flow frequency ($\omega$). We thus obtain $\tau$ from the programmed $\omega$, as a function of T. By plotting the $\log(\tau)$ against $T_g/T$, we then deduced the fragility index m from the slope of the Arrhenius plot using equation (1).

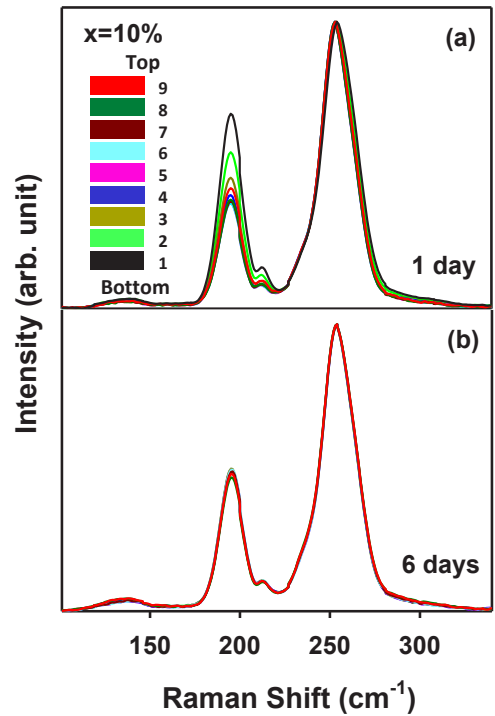

**Figure 3** Raman profiled data of a $Ge_xSe_{100-x}$ melt at x = 10% after (a) $t_R$ = 1 day and (b) $t_R$ = 6 days of melt reaction at 950°C. In (a) there are 9 spectra taken along the length of the melt column, while in (b) these 9 spectra become indistinguishable as melts homogenize. Profiled Spectra taken at other reaction times are not shown here. Variations in line-shape reflect changes in melt stoichiometry along the length of the melt column. See ref.[7] for details. The excitation radiation was 1064nm and laser spot size 50μm.

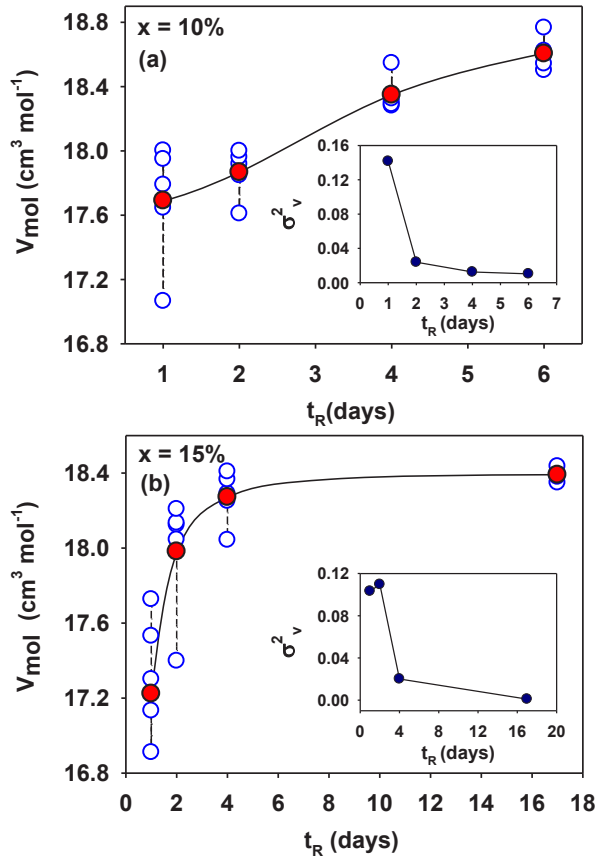

**Figure 4** Molar volumes of melt-quench glasses at (a) $x = 10\%$, and at (b) $x = 15\%$ of Ge illustrating the variance in and saturation of $V_m(t_R)$ as melts homogenize. The filled data point is the mean $V_m$ value. Note that the spread of $V_m$ data points at a given $t_R$, the variance ($\sigma^2_v$), is large at low $t_R$ but it steadily decreases with $t_R$ as melts homogenize, as graphed in the two insets.

**3.1 Molar volumes** A melt at $x = 10\%$, after $t_R = 1d$ is quite heterogeneous but after $t_R = 6d$ (Fig.3) of reaction becomes homogeneous as documented by the FT-Raman profiling. In Fig.4a, we show results obtained at $x = 10\%$, and find that at $t_R = 1d$, $V_m$ data show a significant spread across the batch composition. The 5 open circle data points represent results on 5 distinct samples taken from the same 2 gram batch composition. The mean value of $V_m$ is shown by a filled circle (red) data point. The spread in $V_m$ between the 5 data points provides a measure of variance that tracks sample heterogeneity. Note that as the melt is homogenized variance in $V_m(x)$ decreases, and the mean value of $V_m$ steadily increases to saturate at a value of 18.60(4) cm$^3$/mole. A similar pattern is observed at $x = 15\%$ (Fig.4b); as the melt is homogenized, the variance in $V_m$ decreases, and the mean value of $V_m$ increases and saturates at 18.40(4) cm$^3$/mole.

We have projected the mean value of $V_m(t_R)$ at $x = 10\%$ and at $x = 15\%$ on a global plot of $V_m(x)$ in Fig.5. On this plot one can see that the saturation value of $V_m$ at $x = 10\%$ and at $x = 15\%$, when melts are homogeneous, they acquire values that are in excellent agreement to the results on the specially homogenized glasses reported earlier[7, 8]. These data provide an internal consistency check on these $V_m(x)$ results. The low values of $V_m$ that are associated with large variance constitute results that are characteristic of heterogeneous samples.

In Fig.5, the vertical blue panel gives the Intermediate phase, while the horizontal curved band encompasses data on $V_m$ from the reports of Yang et al.[16], Feltz et al.[14] and Avtikyan et al.[15] from Fig.1.

**3.2 Fragility** A 10 mg quantity of the quenched melt, hermetically sealed in Al pans was cooled from $T_g + 20°C$ to room temperature followed by a heating cycle at a scan rate of 1°C/min and with modulation time varied between 60sec to 140sec in the Q2000 modulated DSC system. From these data we extracted the relaxation time $\tau$ as a function of T, and deduced the fragility index from the Arrehenian plot. Melt quenched glasses at $x = 10\%$ were examined as a function of reaction time $t_R$ in the $1d < t_R < 6d$. At each $t_R$, 3 samples were studied, and the results are summarized in Fig.6.

The three open circle data points in Fig. 6 represent the measured fragility index, while the filled circle data point gives the mean value of the fragility index. One can see that the variance in the fragility index is large at $t_R = 1d$, and it steadily decreases as $t_R$ increases to 6d as melts homogenize as monitored by Raman profiling experiments. Although the mean value of the fragility index decreases with $t_R$, it seems to go through a minimum near $t_R = 4d$.

Nevertheless the data clearly reveal melt heterogeneity to decrease as shown from the time evolution of the variance $\sigma^2_m$ in our measurements (inset of Fig.6). The fragility index for the most homogeneous sample at $t_R = 6d$ of m = 25 is in reasonably good accord with the value reported earlier by us for the completely homogenized melts in Fig.2. We shall comment on these results next.

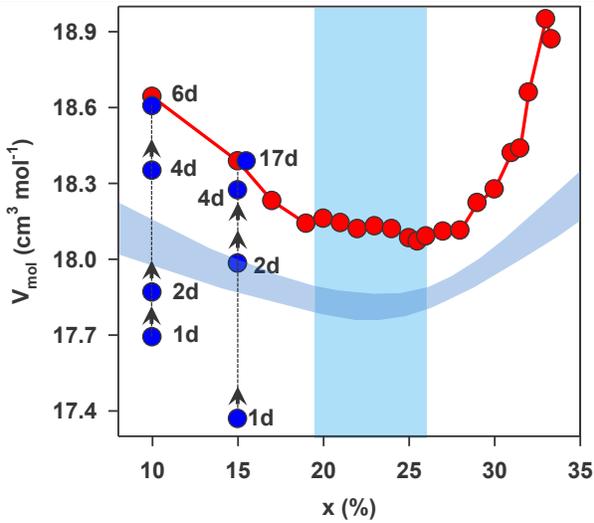
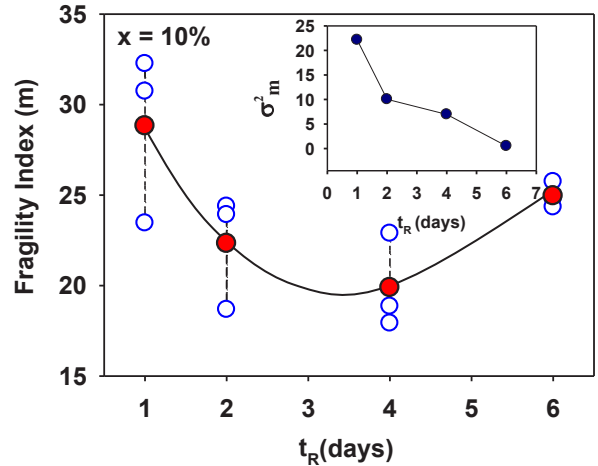

**Figure 5** $V_m$ of the melt-quenched glass at x = 10% and x = 15% from figure 2 are projected on the global $V_m(x)$ variation ob-observed (●)in dry homogeneous $Ge_xSe_{100-x}$ glasses Bhosle et al. ref [12]. Note that heterogeneous glasses have low $V_m$, and as they homogenize Vm increases and saturates at values characteristic of the homogeneous glasses reported earlier by Bhosle et al. See ref.[8]. The shaded vertical panel represents the Intermediate Phase, while the horizontal curved band gives the range of $V_m$ reported in ref[14-16].

**Figure 6** Variation in fragility index m ($t_R$) of $Ge_{10}Se_{90}$ melt as a function of melt reaction time $t_R$ over 6 days as these homogenize. The open circle show results on 3 independent samples. The filled circle is the mean value of m. The inset shows corresponding variance $\sigma^2_m$.

## 4 Discussion

### 4.1 Melt Heterogeneity and interfacial regions

The principal findings of an increase in $V_m$ and a decrease of m-index as melts/glasses are homogenized can be commented upon now. In the early stages of reacting the starting materials, particularly at $t_R < 2d$, the measured $V_m$ are quite low, in fact lower than the broad range of values in the 18.1 cm$^3$/mol < $V_m$ < 18.6 cm$^3$/mol band that is characteristic of homogeneous glasses (Fig.5). For this reason, one cannot merely view the heterogeneous glasses ($t_R > 1d$) to be a mere superposition of homogeneous domains of varying stoichiometry $x_i$ in the 0 < $x_i$ < 33.3% range. There are regions in such heterogeneous melts/glasses that are quite compacted. At rather short reaction times, $t_R < 1d$, it is indeed true that crystalline phases form. However, such phases steadily disappear as melts are reacted longer for $t_R > 1d$ (Fig. 3). These $V_m$ data are suggestive that heterogeneous melts may be viewed as composed of *homogeneous regions* of well-defined stoichiometry "$x_i$" that are separated by *heterogeneous interfacial regions* as schematically illustrated in Fig.7. We view the well-defined *homogeneous regions* to be composed of characteristic local structures (Se$_n$ chain fragments, GeSe$_2$ –Corner-sharing(CS) and Edge-Sharing(ES) tetrahedral units) with well-developed extended range structures, such as fraction of ES/CS fixed by stoichiometry $x_i$ alone, which give rise to the appropriate mode signature in Raman experiment. On the other hand, *interfacial regions* are viewed as regions that connect homogeneous regions of varying stoichiometry. They are largely composed of the same local structures as the *homogeneous* regions but could have Ge-rich local structures and broken bonds, but with the important difference that *extended range structures are not developed*. We view interfacial regions to possess low molar volumes and high fragility index, features that we associate with *absence* of extended range structures. As melts homogenize upon increased $t_R$, homogeneous regions grow by reconstructing with interfacial ones as schematically illustrated in Fig.7a and b, and the process saturates as $V_m$ increases (Fig.4 )and m decreases (Fig 6) to acquire values characteristic of the completely homogeneous melts/glasses.

### 4.2 Slow kinetics of melt homogenization

Why are the kinetics of melt homogenization slow? In the early phase ( ~ 1 day) of reacting elemental Ge with Se, melts of increasing Se-stoichiometry form along the length of the column starting from the tube bottom up as noted in the present Raman profiling experiments (Fig. 2) and also earlier work[7, 8, 29]. The density of liquid Ge exceeds that of liquid Se, resulting in melts towards the tube bottom to be Ge-rich. But as $t_R$ increases concentration gradients dissipate as Ge(Se) atoms diffuse up (down) the melt column. Fragility data on homogeneous melts unequivocally shows that melts in the composition range, 20% < x < 26%, are strong, while those in the narrow range, 21.5% < x < 23%, to be super-strong (Fig. 2), i.e. possess a fragility of 14.8(5) that is even lower than that of silica[4]. These melts have a viscosity at the reaction T (950°C) that exceed those of fragile melt compositions outside the 20% < x < 26% window by a factor of 40 or more. Since melt diffusivities are inversely proportional to viscosity through the Eyring relation [30]

$$D \propto 1/\eta \quad (2)$$

one expects D to be about two orders of magnitude lower for the super-strong melts than for the fragile ones. These wide differences in diffusivities slow down the kinetics of melt homogenization. One expects the time needed to homogenize a melt composition to be batch size or diffusion-length dependent with larger melts taking longer to homogenize than smaller ones[7]. Experiments, indeed, confirm that prediction as discussed in detail elsewhere[7, 8]. Since the underlying process is diffusive in character one does not expect convective mixing of melts alone to dramatically alter the kinetics of melt-homogenization. Convective mixing such as rocking of melts will assist in overcoming gravitation induced segregation of liquids in the early stages of reaction[6,7], but as these large scale segregation effects dissipate, ultimately it is the diffusive processes that control atomic scale mixing of melts. Diffusion in chalcogenide liquids has been recently investigated[31] and it has been found that at 820°C, the diffusion constant of $GeSe_2$ is about $D = 2 \times 10^{-10}$ m$^2$/s, in agreement with an estimate combining the Eyring equation for liquids and measurements of viscosity. Using the definition (Einstein relation) of the mean square displacement $<r^2(t)> = 6tD$, one can thus estimate that a particle will diffuse through a length of $<r^2(t)>^{1/2}$ = 3cm after $t = t_R \sim 8.5$ days. The mean square displacement in the diffusive régime is always proportional to time. Clearly then, a reaction time, $t_R$, an order of magnitude less will not permit Ge and Se atoms to fully diffuse across the melt. Additionally, these $t_R$ will have to obviously increase if the reaction T is lowered (D decreasing with T) or if the batch size is increased.

### 4.3 Broader Implications

#### 4.3.1 Generality of the intermediate phase in network glasses
The introduction of Rigidity Theory to understanding network glasses since the 1980s has stimulated a large body of theoretical and experimental work[32]. The crucial role of network topology in systematically altering physical properties of network glasses has led to the recognition of two underlying elastic phase transitions[33], a *rigidity* transition followed by a *stress* transition. The nature of these transitions and their structural manifestations continues to be a subject of current interest. These transitions have now been observed in different types of material systems including heavy metal oxides[20] and modified oxides[34], in addition to chalcogenides[3, 12, 17, 35]. They have been also observed in realistic molecular simulations[36, 37]. Given these new findings, investigations of these elastic phase transitions in glassy solids will hold the key to understanding the phenomenon of self-organization and the rather special physical properties of the phase formed between these two transitions[38].

#### 4.3.2 Melt/glass heterogeneity and denial of the intermediate phase
Melt-quenching as a method to synthesize bulk glasses is deceptively simple. It has been used since the inception of the field of glass science more than 80 years ago. In covalently bonded glass forming systems, melts undergo "slow" homogenization. The popular belief that by suitably reacting starting materials at 200°C to 300°C above their melting temperatures for 24 hours with a continuous rocking regardless of batch sizes, one could achieve melt homogeneity appears not to be supported by experiments[12]. Slow homogenization of covalent glassy melts has had the unfortunate consequence that physical properties reported by various groups on the same material systems display wide variations, as illustrated here for the case of molar volumes and fragility in the Ge-Se binary.

The present finding of an increase in $V_m$ and a decrease in fragility m-index (Fig.4) of Ge-Se melts/glasses as these are steadily homogenized clearly demonstrates that some of the earlier work on these glasses (Fig.1 and 4), particularly those that possess a low $V_m(x)$ must come from specimen that are intrinsically *heterogeneous* by virtue of synthesis. A perusal of Fig.1 suggests that the results of Feltz et al.[14], Avtikyan et al.[15] and Yang et al.[16] display $V_m(x)$ trends that largely reside in the 17.8 – 18.1 cm$^3$/mole range across a wide range of Ge content. This range overlaps with values we observe in our present glasses that were reacted typically for $t_R$ < 2d (Fig.5), which we know from Raman profiling data to be heterogeneous. In the work of Yang et al[16], the authors synthesized 20 to 25 gm batch compositions[39] and reacted the elements at 700° C for 12 h in a rocking furnace. These conditions of synthesis used by Yang et al.[16], we believe, has led to heterogeneous glasses. And the diphasic model[40] of these glasses proposed from $^{77}$Se NMR has substantial fraction of the signal coming from interfacial regions rather than the homogeneous ones. In sharp contrast, the $V_m(x)$ trends reported by Mahadevan et al.[13] that almost straggle the results of Bhosle et al. (Fig.1), are on glass samples that appear reasonably homogeneous.

Chalcogenides are fascinating materials and display a richness of physical phenomenon. However, to unravel these phenomena, particularly the percolative stress- and rigidity- elastic phase transitions in covalent glasses such as the present Ge-Se binary, the need for homogeneous glass is *paramount*. In our earliest work[41] on the subject, these transitions were smeared because glasses were not as homogeneous. We came to recognize this to be the case more recently when these elastic phase transitions became rather *abrupt* in composition in the specially synthesized homogenized melts/glasses[12]. The power of FT-Raman profiling[29] method in monitoring the heterogeneity of melts in this context cannot be overemphasized. The same technique has now afforded us a means to synthesize melts/glass of controlled heterogeneity in the present investigations. The present findings also rule out reported denials [40, 42] of the double percolative transition, rigidity and stress, given that the associated demonstrations have been made from glasses that are obviously heterogeneous.

The experimental challenges of chalcogenides to establish the *intrinsic compositional variation of physical properties* requires that they not only be homogeneous but also dry[7, 8]. We discussed the issue of dryness earlier[8]. It would be useful to mention that in Fig.1 the two blue triangle data points are on samples obtained by vacuum sealing finely powdered Ge-Se mixtures left in the laboratory ambient environment for 24 hours. These glasses picked up water to lower molar volumes enough to be directly measured. Indeed, presence of water traces in melts speeds up the kinetics of melt homogenization by fragmenting Se-chains to assist formation of the crosslinked backbone[7, 8], however such glasses also possess a reduced $V_m$ (Fig.1). These "*wet but homogeneous*" glasses display physical properties, such as $T_g$, enthalpy of relaxation at $T_g$, Raman scattering, that differ distinctly from their "*dry and homogeneous*" counterparts as discussed elsewhere[7, 8].

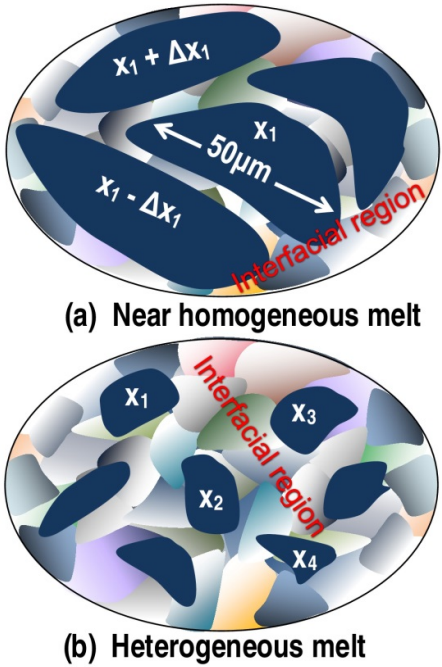

**Figure 7** Schematic of melt homogenization process of present Ge-Se chalcogenides showing (a) growth of homogeneous regions (dark blue) of well-defined melt stoichiometry (x) (b) at the expense of interfacial regions (multicolored slabs). In a heterogeneous melt, regions of varying stoichiometry, $x_1$, $x_2$, $x_3$, occur, but upon homogenization, a unique melt composition $x_1$ persists across the batch composition.

Many sophisticated physical measurements on glasses that probe crucial aspects of local and intermediate range structure require large glass samples, and include neutron scattering, multi-dimensional NMR, bulk elastic constants, and T-dependent viscosity measurements. In most of these reports often little attention is paid to characterization of glass samples by other methods. A simple $T_g$ measurement that uses 20 mg of a 20 gram batch composition samples 0.1% of the glass, and can hardly be a sufficient representation of a glass sample homogeneity.

On the other hand, volumetric measurements can make use of large samples, and they are not only straightforward and inexpensive but also can be used to provide the variance across a batch composition to directly establish glass heterogeneity as demonstrated in the present work.

**4.3.3 Extending notion of network rigidity to liquids** The present findings on glasses are reminiscent of structure related anomalies noted in densified liquids from Molecular Dynamic simulations. For example, in water[43], silica[44, 45] and $BeF_2$, molar volume variations have been reported from simulated equations of state. In these densified tetrahedral liquids one finds molar volumes to display minima as a function of applied external pressure. These are related to anomalies in diffusion constant, changes in orientational and translational order parameters suggesting a coordination number increase from tetrahedral to octahedral with an attendant configurational entropy increase[45].

Recently, it has been shown[37, 46] that such anomalies in the liquid state are manifestations of the stress-free nature of the system which adapts under increasing pressure by releasing some bond-bending interactions in order to accommodate increased stress due to a coordination number increase. Trends showing extrema in different structural and dynamic quantities allow defining a window as a function of applied external pressure, which has striking similarities to the window found in the present chalcogenides glasses as a function of composition at ambient pressure. In this stress-free pressure window[37], activation energies $E_a$ for viscosity and diffusivity are found to display minima, which point to the configurational entropy of the intermediate phase liquid to be a maximum (Adam-Gibbs relationship). These results are fully in line with the observed minima of fragility and $E_a$ for relaxation of Ge-Se melts (Ref. 4 ) and in the non-reversing enthalpy of relaxation at $T_g$ of glasses. The vanishing of the non-reversing enthalpy is direct evidence that glass compositions in the Intermediate Phase [7,8], behave "liquid-like" and possess a *high* configurational entropy.

**5 Conclusions** Changes in physical properties including molar volumes, fragility, and Raman vibrational density of states of 2 gram sized $Ge_xSe_{100-x}$ batches at x = 10% and 15% are closely followed as melts/glasses are steadily homogenized. Molar volumes, $V_m$ increase as batches homogenize to saturate at values characteristic of homogeneous glasses. Fragility index, m, steadily decreases as batches are homogenized, and saturate at values characteristic of homogeneous melts. In both cases, the *variance* in $V_m$ and m steadily decreases as melts/glasses are homogenized. These findings demonstrate that to establish the intrinsic compositional trends in physical properties of non-stoichiometric chalcogenides glasses, it is paramount to homogenize batch compositions. Fundamentally, chalcogenide melts will undergo slow homogenization because of the superstrong nature of select melt compositions in the Intermediate Phase. The rigidity and stress- elastic phase transitions are smeared in heterogeneous glasses but become rather abrupt in homogeneous ones.

It is most unfortunate that aspects of sample synthesis have been overlooked during the recent debate challenging the existence of the double transition and the intermediate phase. In the present contribution we emphasize the crucial importance of melt homogenization for the detection of the subtle elastic changes at play over small compositional changes.

**Acknowledgements** It is a pleasure to acknowledge discussions with J.C. Phillips during the course of this work. This work is supported by NSF grant DMR 08-53957 and ANR Grant No. 09-BLAN-0109-01. MM acknowledges financial support from International Materials Institute, and from the French-American Fulbright Commission.

**Rajat Bhageria is a rising senior at Sycamore High School, Cincinnati, OH. He has been involved in research at University of Cincinnati working in the laboratory of Dr. P. Boolchand since fall 2011.

**References**

[1] J. C. Phillips, Phys. Rev. E **80** 051916 (2009).
[2] R. Brückner, J. of Non-Cryst. Solids **5** 123-175 (1970).
[3] K. Gunasekera, P. Boolchand and M. Micoulaut, The Journal of Physical Chemistry B (Accepted) (2013).
[4] K. Gunasekera, S. Bhosle, P. Boolchand and M. Micoulaut, (In preparation) (2013).
[5] C. Bourgel, M. Micoulaut, M. Malki and P. Simon, Phys. Rev. B **79** 024201 (2009).
[6] P. Boolchand, J. C. Mauro and J. C. Phillips, Physics Today **66** 10-11 (2013).
[7] S. Bhosle, K.Gunasekera, P. Boolchand and M. Micoulaut, Intl. J. App. Glass. Sci. **3** 189-204 (2012).
[8] S. Bhosle, K. Gunasekera and P. Boolchand, Intl. J. App. Glass. Sci. **3** 205-220 (2012).
[9] J. C. Phillips and M. F. Thorpe, Solid State Commun. **53** 699-702 (1985).
[10] R. Kerner and J. C. Phillips, Solid State Commun. **117** 47-51 (2000).
[11] R. C. Welch, J. R. Smith, M. Potuzak, X. Guo, B. F. Bowden, T. J. Kiczenski, D. C. Allan, E. A. King, A. J. Ellison and J. C. Mauro, Phys. Rev. Lett. **110** 265901 (2013).
[12] S. Bhosle, K. Gunasekera, P. Chen, P. Boolchand, M. Micoulaut and C. Massabrio, Solid State Commun. **151** 1851-1855 (2011).
[13] S. Mahadevan, A. Giridhar and A. K. Singh, Indian Journal of Pure & Applied Physics **33** 643-652 (1995).
[14] A. Feltz, H. Aust and Blayer, J. of Non-Cryst. Solids **55** 179-190 (1983).
[15] G. B. Avetikyan and L. A. Baidakov, Izv. Akad. Nauk SSSR, Neorg. Mater. **8** 1489-1490 (1972).
[16] G. Yang, Y. Gueguen, J.-C. Sangleboeuf, T. Rouxel, C. Boussard-Plédel, J. Troles, P. Lucas and B. Bureau, J. of Non-Cryst. Solids (In press)
[17] S. Chakravarty, D. G. Georgiev, P. Boolchand and M. Micoulaut, Journal of Physics-Condensed Matter **17** L1-L7 (2005).
[18] S. Chakraborty and P. Boolchand, (In preparation) (2013).
[19] K. Rompicharla, D. I. Novita, P. Chen, P. Boolchand, M. Micoulaut and W. Huff, Journal of Physics: Condensed Matter **20** 202101 (2008).
[20] S. Chakraborty, presented at the APS March Meeting, Dallas, TX, 2011 (unpublished).
[21] M. W. Deem and J. C. Phillips, (2013) arXiv:1308.5718
[22] X. Hu, S. C. Ng, Y. P. Feng and Y. Li, Phys. Rev. B **64** 172201 (2001).
[23] H. N. Ritland, J. of the Am. Ceram.Society **37** 370-377 (1954).
[24] M. Tatsumisago, B. L. Halfpap, J. L. Green, S. M. Lindsay and C. A. Angell, Phys. Rev. Lett. **64** 1549 (1990).
[25] M. Micoulaut, J. Phys.: Condens. Matter **22** 1-7 (2010).
[26] U. Senapati and A. K. Varshneya, J. of Non-Cryst. Solids **197** 210-218 (1996).
[27] S. Stolen, T. Grande and H.-B. Johnsen, Physical Chemistry Chemical Physics **4** 3396-3399 (2002).
[28] L. C. Thomas, *Modulated DSC Technology (MSDC-2006)*. (T.A. Instruments, Inc (www.tainstruments.com), New Castle, DE, 2006).
[29] P. Boolchand, K. Gunasekera and S. Bhosle, Physica Status Solidi B 1-6 (2012).
[30] J. P. Hansen and I. R. McDonald, *Theory of Simple Liquids*. (Academic Press, 2006).
[31] M. Micoulaut and C. Massobrio, Journal of Optoelectronics and Advanced Materials **11** 1907-1914 (2009).
[32] P. Boolchand, G. Lucovsky, J. C. Phillips and M. F. Thorpe, Philosophical Magazine **85** 3823-3838 (2005).
[33] M. F. Thorpe, D. J. Jacobs, M. V. Chubynsky and J. C. Phillips, J. of Non-Cryst. Solids **266** 859-866 (2000).
[34] K. Vignarooban, P. Boolchand, M. Micoulaut and M. Malki, presented at the APS March Meeting, Boston, MA, 2012 (unpublished).
[35] D. Selvanathan, W. J. Bresser, P. Boolchand and B. Goodman, Solid State Commun. **111** 619-624 (1999).
[36] M. Bauchy, M. Micoulaut, M. Boero and C. Massobrio, Phys. Rev. Lett. **110** 165501 (2013).
[37] M. Bauchy and M. Micoulaut, Phys. Rev. Lett. **110** 095501 (2013).
[38] P. Boolchand, D. G. Georgiev and B. Goodman, Journal of Optoelectronics and Advanced Materials **3** 703-720 (2001).
[39] B. Bureau, J. Troles, M. Le Floch, F. Smektala and J. Lucas, J. of Non-Cryst. Solids **326** 58-63 (2003).
[40] P. Lucas, E. A. King, O. Gulbiten, J. L. Yarger, E. Soignard and B. Bureau, Phys. Rev. B **80** 214114 (2009).
[41] X. W. Feng, W. J. Bresser and P. Boolchand, Phys. Rev. Lett. **78** 4422-4425 (1997).
[42] G. Yang, B. Bureau, T. Rouxel, Y. Gueguen, O. Gulbiten, C. Roiland, E. Soignard, J. L. Yarger, J. Troles, J.-C. Sangleboeuf and P. Lucas, Phys. Rev. B **82** 195206 (2010).

[43] J. R. Errington and P. G. Debenedetti, Nature **409** 318-321 (2001).
[44] M. S. Shell, P. G. Debenedetti and A. Z. Panagiotopoulos, Phys. Rev. E **66** 011202 (2002).
[45] B. S. Jabes, M. Agarwal and C. Chakravarty, J. of Chem. Phys. **132** 234507-234512 (2010).
[46] M. Micoulaut and M. Bauchy, physica status solidi (b) **250** 976-982 (2013).